\documentclass[preprint,12pt]{elsarticle}

\usepackage{listings}
\usepackage{amssymb}

\usepackage{verbatim}
\usepackage{alltt}

 \usepackage{lineno}

\usepackage{rotating}

\newcounter{bla}

\journal{Computer Physics Communications}

\begin{document}

\begin{frontmatter}



\title{gSeaGen: the KM3NeT GENIE-based code for neutrino telescopes}

\cortext[cor]{corresponding author}

\author[a]{S.~Aiello}
\author[bb,b]{A.~Albert}
\author[c]{S.~Alves~Garre}
\author[d]{Z.~Aly}
\author[e]{F.~Ameli}
\author[f]{M.~Andre}
\author[g]{G.~Androulakis}
\author[h]{M.~Anghinolfi}
\author[i]{M.~Anguita}
\author[j]{G.~Anton}
% ...
\address[a]{INFN, Sezione di Catania, Via Santa Sofia 64, Catania, 95123 Italy}
\address[b]{IN2P3, IPHC, 23 rue du Loess, Strasbourg, 67037 France}
\address[c]{IFIC - Instituto de F{\'\i}sica Corpuscular (CSIC - Universitat de Val{\`e}ncia), c/Catedr{\'a}tico Jos{\'e} Beltr{\'a}n, 2, 46980 Paterna, Valencia, Spain}
\address[d]{Aix~Marseille~Univ,~CNRS/IN2P3,~CPPM,~Marseille,~France}
\address[e]{INFN, Sezione di Roma, Piazzale Aldo Moro 2, Roma, 00185 Italy}
\address[f]{Universitat Polit{\`e}cnica de Catalunya, Laboratori d'Aplicacions Bioac{\'u}stiques, Centre Tecnol{\`o}gic de Vilanova i la Geltr{\'u}, Avda. Rambla Exposici{\'o}, s/n, Vilanova i la Geltr{\'u}, 08800 Spain}
\address[g]{NCSR Demokritos, Institute of Nuclear and Particle Physics, Ag. Paraskevi Attikis, Athens, 15310 Greece}
\address[h]{INFN, Sezione di Genova, Via Dodecaneso 33, Genova, 16146 Italy}
\address[i]{University of Granada, Dept.~of Computer Architecture and Technology/CITIC, 18071 Granada, Spain}
\address[j]{Friedrich-Alexander-Universit{\"a}t Erlangen-N{\"u}rnberg, Erlangen Centre for Astroparticle Physics, Erwin-Rommel-Stra{\ss}e 1, 91058 Erlangen, Germany}
% ...
\address[bb]{Universit{\'e} de Strasbourg, Universit{\'e} de Haute Alsace, GRPHE, 34, Rue du Grillenbreit, Colmar, 68008 France}
% ...

\begin{abstract}
The gSeaGen code is a GENIE-based application developed to efficiently generate high statistics samples of events, induced by neutrino interactions, detectable in a neutrino telescope. The gSeaGen code is able to generate events induced by all neutrino flavours, considering topological differences between track-type and shower-like events. Neutrino interactions are simulated taking into account the density and the composition of the media surrounding the detector. The main features of gSeaGen are presented together with some examples of its application within the KM3NeT project.
\end{abstract}

\begin{keyword}
Astroparticle Physics \sep High Energy Neutrinos \sep Monte Carlo Event Generator \sep Neutrino Telescopes \sep Neutrino Oscillations \sep KM3NeT \sep GENIE. 
\end{keyword}

\end{frontmatter}


\begin{small}
\noindent
{\em Manuscript Title:} gSeaGen: a GENIE-based code for neutrino telescopes                                       \\
\\
{\em Authors: } The KM3NeT Collaboration (corresponding author C. Distefano)                                       \\
\\
{\em Program Title:} gSeaGen                                          \\
\\
{\em Journal Reference:}                                      \\
\\
{\em Catalogue identifier:}                                   \\
\\
{\em Licensing provisions:} GPLv3                                  \\
\\
{\em Programming language:} C++                                 \\
\\
{\em Computer:} CC-IN2P3 Linux computer farm: platform CentOS 7.6.1810, architecture: x86\_64 (see https://cc.in2p3.fr/).                                               \\
\\
{\em Operating system:} Linux and Unix-like operating systems.                                     \\
\\
{\em RAM:} 550 Mbytes                                              \\
\\
{\em Number of processors used:} one                             \\
\\
{\em Supplementary material:}                                 \\
\\  
{\em Keywords:} Astroparticle Physics, High Energy Neutrinos, Monte Carlo Event Generator, Neutrino Telescopes, Neutrino Oscillations, KM3NeT, GENIE. \\
\\
{\em Classification:} 1.1 Cosmic Rays,  11.1 General, High Energy Physics and Computing, 11.3 Cascade and Shower Simulation, 11.7 Detector Design and Simulation. \\ 
\\
{\em External routines/libraries:} GENIE \cite{1} and its external dependencies. Linkable to MUSIC \cite{2} and PROPOSAL \cite{3}.  \\
\\
{\em Subprograms used:}                                       \\
\\
{\em Catalogue identifier of previous version:}*              \\
\\
{\em Journal reference of previous version:}*                  \\
\\
{\em Does the new version supersede the previous version?:}*   \\
\\
{\em Nature of problem:}
Development of a code to generate detectable events in neutrino telescopes, using modern and maintained neutrino interaction simulation libraries which include the state-of-the-art physics models.
The default application is the simulation of neutrino interactions within KM3NeT \cite{4}.\\
\\
{\em Solution method:}
Neutrino interactions are simulated using GENIE, a modern framework for Monte Carlo event generators. The GENIE framework, used by nearly all modern neutrino experiments, is considered as a reference code within the neutrino community.\\   
\\
{\em Reasons for the new version:}*\\
   \\
{\em Summary of revisions:}*\\
   \\
{\em Restrictions:} Simulation of neutrinos with energy lower than 5 TeV, due to the present GENIE valid energy range.\\
   \\
{\em Unusual features:}\\
   \\
{\em Additional comments:} The code was tested with GENIE version 2.12.10 and it is linkable with release series 3. \\
   \\
{\em Running time:} 3 ms per generated muon neutrino with energy of 1 TeV, using a 1.3 GHz CPU.\\

* Items marked with an asterisk are only required for new versions
of programs previously published in the CPC Program Library.\\
\end{small}


\section{Introduction}
\label{sec:intro}

Monte Carlo simulations play an important role in the data analysis of neutrino telescopes. 
Simulations are used to design reconstruction algorithms for neutrino events, as well as to estimate cosmic and atmospheric signals in various physics analyses.
GENIE \cite{Andreopoulos:2009rq} is a neutrino event generator focused mainly on the low-energy range ($\le$ 5 GeV) and presently valid up to 5 TeV. It is currently used by many experiments working in the neutrino oscillation field e.g.\ \cite{t2k, nova, microboone, minerva, dune, sbn}. 
The project's goal is the development of a reference neutrino interaction Monte Carlo simulation tool whose validity extends to PeV scales. 

The gSeaGen code simulates events induced by all flavours of neutrinos and detectable by a neutrino telescope, using GENIE to simulate the neutrino interactions.  
In particular, it allows electron, muon and tau neutrino events to be generated. Depending on the steering parameters it can produce events from multiple (anti)neutrino flavours within one execution of the program (e.g. $\nu_\mu$ + $\bar{\nu}_\mu$); simulate charged current (CC) and/or neutral current (NC) interactions; and simulate neutrinos arriving isotropically or from point or extended sources. The code was developed for KM3NeT \cite{km3net-loi}. Although applicable both to underwater and under-ice neutrino detectors, the code terminology, as well as the description presented in this paper, follow those of underwater detectors.

\section{The coordinate system}

\begin{figure}[h]
\centering
\includegraphics[scale=0.3]{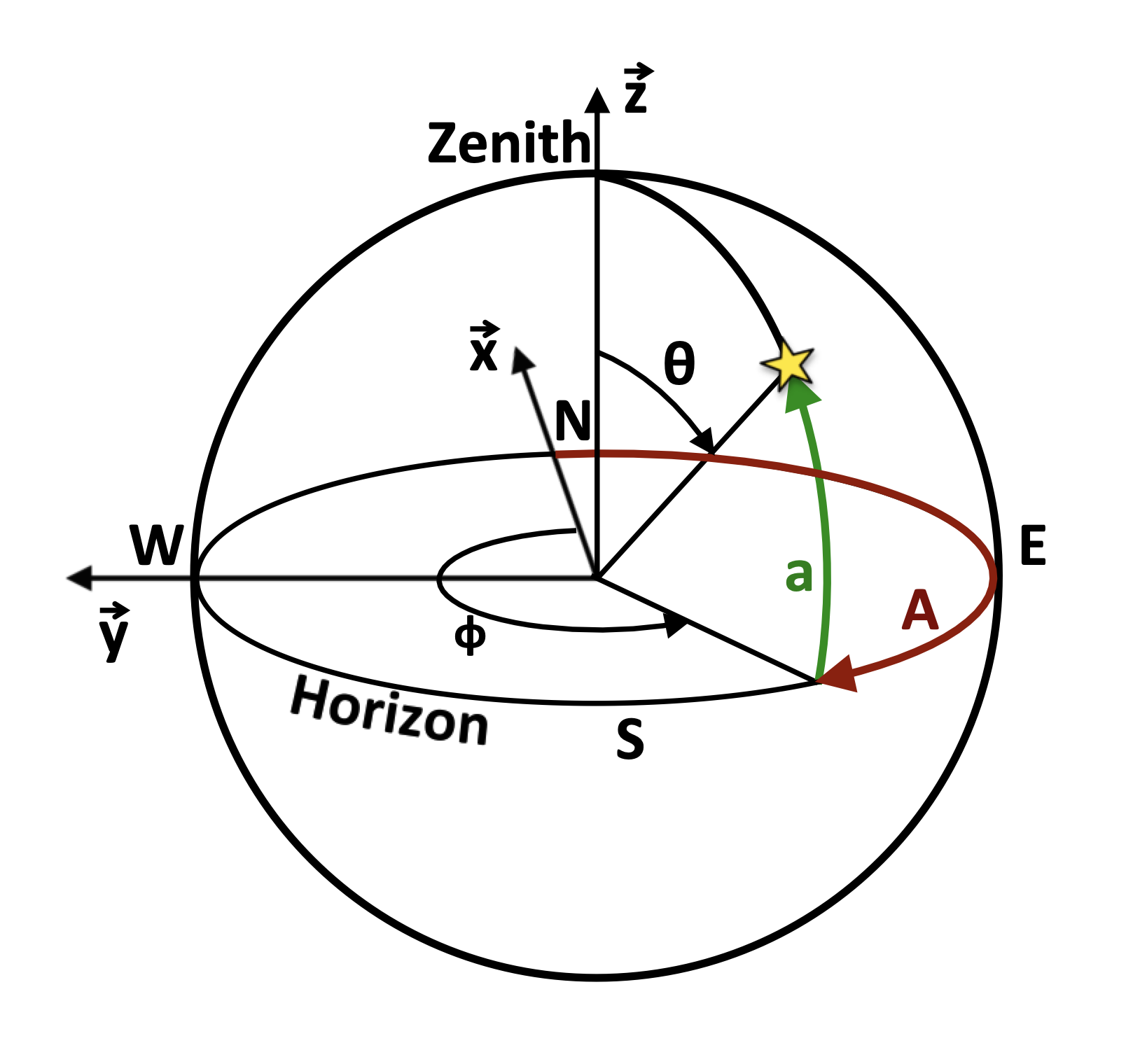}
\caption{The Cartesian and the associated astronomical horizontal systems as defined in gSeaGen. The relation between the source altitude $a$ and azimuth $A$ coordinates and the angles $\theta$ and $\phi$, defining the arrival direction of the emitted neutrinos is shown.}
\label{fig:coord}       
\end{figure}

The coordinates in gSeaGen are Cartesian with the positive $z$-axis pointing to the local zenith, the positive $x$-axis pointing north and the positive $y$-axis as needed to make a right-handed coordinate system. By default, the origin is defined as being the water/rock boundary.
To define astrophysical source positions, an associated horizontal coordinate system is defined, with the azimuth measured eastward from the North, as shown in Fig.\ref{fig:coord}.
The relation between the source altitude $a$ and azimuth $A$ coordinates and the angles $\theta$ and $\phi$, defining the arrival direction of the emitted neutrinos, is then:

\begin{equation}
\left\{
\begin{array}{lll}
\theta  & = & 90^\circ-a\\ 
        &   &  \\
\phi & = & 360^\circ-A  \\
\end{array}.
\right. 
\label{eq:hor2det} 
\end{equation}

\section{The detector active volume}
\label{sec:can}

\begin{figure}[h]
\centering
\includegraphics[scale=0.4]{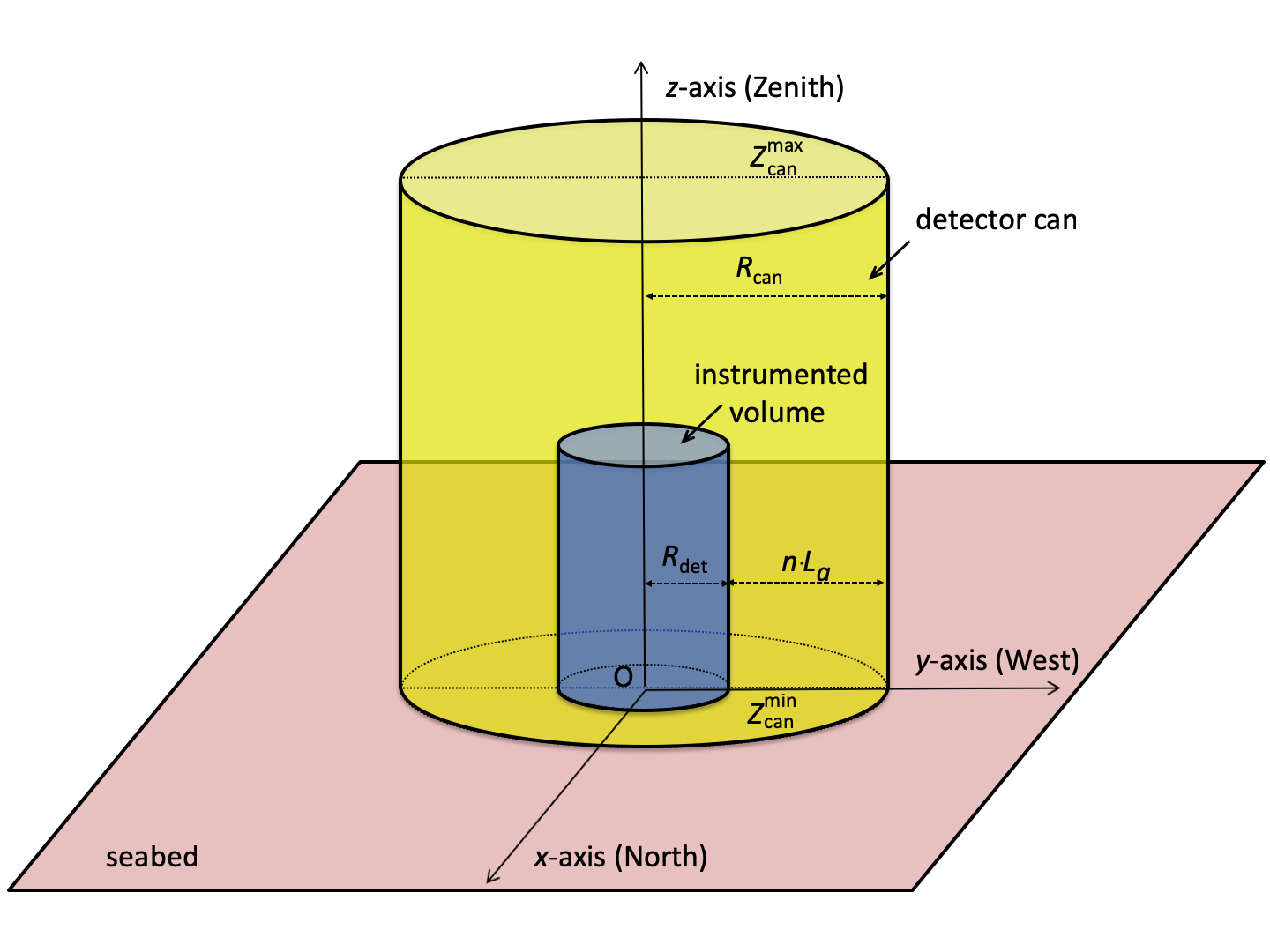}
\caption{Definition of the detector can.}
\label{fig:can}       
\end{figure}

The detector active volume is defined as a cylinder, the {\it can}, having the bottom bounded at the seabed and surrounding the instrumented volume as sketched in Fig. \ref{fig:can}.
The user can set the can defining the $z$-level of the can bottom $Z^{\rm min}_{\rm can}$, the $z$-level of the can top $Z^{\rm max}_{\rm can}$ and the can radius $R_{\rm can}$.
It is possible to define $Z^{\rm min}_{\rm can}\ne0$. In this case, the coordinate system origin is translated along the $z$-axis in order to bound the can to the bottom at the seabed. With this choice, the user can  shift the origin along the $z$-axis from the default position set at $z=0$.

Presently, the bottom of the can is bounded at the seabed; this constraint will be overcome in the next versions.
However, detectors not lying on the rock, as in the case of under-ice telescopes, can be simulated by enlarging the can below the detector. 

gSeaGen can also generate a can by automatically reading a file containing the detector geometry. In this case the code defines a cylinder representing the detector's instrumented volume. The can is then built by enlarging the instrumented volume by $n$  times the light absorption length, $L_a$, input by the user. Presently only ANTARES and KM3NeT geometry files are accepted by gSeaGen.

The can represents the detector horizon to the light emitted by the neutrino-induced particles. All the events leading to Cherenkov radiation in the can, i.e. events interacting inside the can or producing a muon reaching the can surface, are stored in the output file. The particle tracking inside the can and the Cherenkov light emission are performed in the next steps of the simulation chain by independent programs to study the detector response.  

\section{The target media}
\label{sec:media}

Four different neutrino target media are defined: {\it SeaWater}, {\it Rock}, {\it Mantle} and {\it Core}. The first two are used to define the interaction volume (see Sec. \ref{sec:vol}), while the others enter in the calculation of the transmission probability of neutrinos through the Earth (see Sec. \ref{sec:wgt}). 
The default compositions for each target medium are listed in Tab. \ref{tab:media}. 
The densities for SeaWater and Rock are set to $\rho_{s}=1.04$ g/cm$^3$ and $\rho_{r}=2.65$ g/cm$^3$ respectively. 
For Mantle and Core media, the density profile provided by the Preliminary Reference Earth Model (PREM) \cite{PREM} is used.

The user can change the composition of all target media and the density for SeaWater and Rock, e.g.\ to study the systematic uncertainties due to medium compositions and also consider under-ice detectors.

\begin{table} 
\caption{Default compositions for target media defined in gSeaGen.} 
\begin{center} 
\begin{small} 
\begin{tabular}{c|c|c|c} 
\hline \hline 
\multicolumn{4}{c}{SeaWater}\\ 
\hline
Element  & Percent  & Element & Percent \\
\hline
O$^{16}$ 	 &  85.84  & S$^{32}$   & 0.091		\\
H$^{1}$ 	 & 10.82   & Ca$^{40}$ & 0.04		\\
Cl$^{35}$  & 1.94     & K$^{39}$   & 0.04		\\
Na$^{23}$ & 1.08 	& Br$^{80}$ & 0.0067	\\
Mg$^{24}$ & 0.1292 & C$^{12}$  & 0.0028	\\ 
\hline \hline 
\multicolumn{4}{c}{Rock}\\ 
\hline
Element  & Percent  & Element & Percent \\
\hline
O$^{16}$ 	 &  46.3  & Na$^{23}$    & 2.36	\\ 
Si$^{28}$ 	 &  28.2  & Mg$^{24}$    & 2.33	\\ 
Al$^{27}$ 	 &  8.23  & K$^{39}$ 	     & 2.09	\\ 
Fe$^{56}$ &  5.63  & Ti$^{48}$     & 0.57	\\ 
Ca$^{40}$ &  4.15  & H$^{1}$       & 0.14 	\\ 
\hline \hline 
\multicolumn{4}{c}{Mantle}\\ 
\hline
Element  & Percent  & Element & Percent \\
\hline
O$^{16}$ 	 &  45.22 	 &  Fe$^{56}$ 	 &  5.97 \\
Mg$^{24}$ &  22.83 	 &  Al$^{27}$ 	 &  2.25 \\
Si$^{28}$ 	 &  21.49 	 &  Ca$^{40}$ 	 &  2.24 \\ 
\hline \hline 
\multicolumn{4}{c}{Core}\\ 
\hline
Element  & Percent  & Element & Percent \\
\hline
Fe$^{56}$ 	& 90.0 &	Ni$^{58}$ 	& 10.0 \\
\hline \hline 
\end{tabular} 
\end{small} 
\end{center} 
\label{tab:media}
\end{table}

\section{The neutrino interaction volume}
\label{sec:vol} 

The {\it interaction volume} is the volume surrounding the detector in which a neutrino interaction can produce detectable particles. 
It is defined at runtime and is used to simulate the interaction of all generated neutrinos.
The interaction volume is built using the TGeoManager ROOT class \cite{ANTCHEVA20092499} and is used to configure the GENIE geometry driver. 
The interaction volume is defined as a cylinder surrounding the can and is made of sea water and rock, whose composition and density are defined using the target media SeaWater and Rock, respectively.

The size and nature of the interaction volume depends on the topology of the generated neutrino events.
In the case of electron neutrino CC interactions and all neutrino flavour NC interactions, a particle shower is produced. 
Such events, here referred to as {\it shower-like events}, may be detected only if the neutrino interacts inside the light-sensitive volume. In this case the interaction volume is defined as a cylinder coincident with the can and entirely made of seawater. It is possible to extend the interaction volume below the can by adding a layer of rock with the option \texttt{-bedrock} (see Sec. \ref{sec:usage}).

\begin{figure}[ht]
\centering
\includegraphics[scale=0.4]{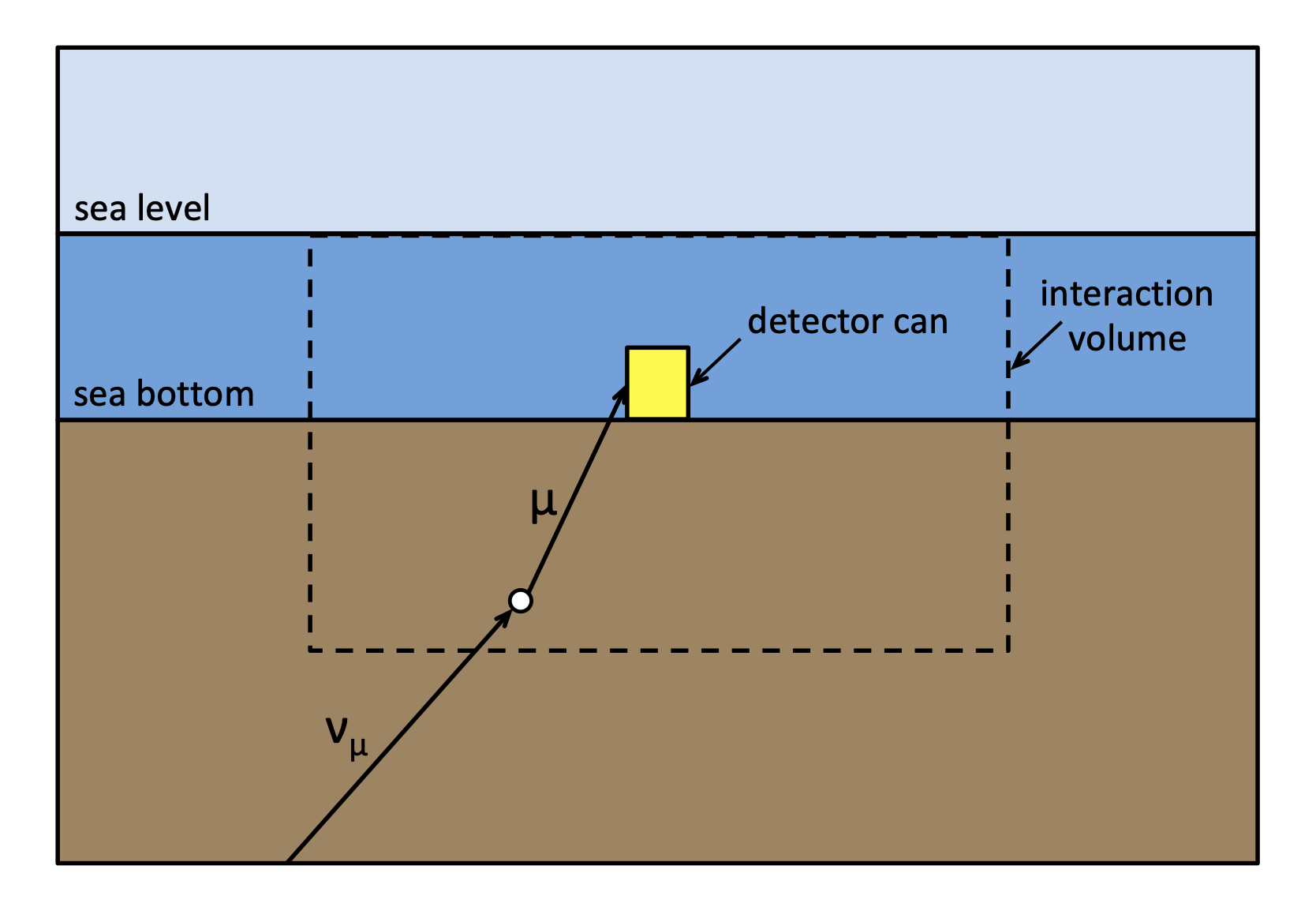}
\caption{Schematic view of the interaction volume for track-like events.}
\label{fig:volume}       
\end{figure}

Muons produced in the neutrino interaction may be detected also if the interaction vertex is outside the can. This happens in the case of muon neutrinos interacting in CC, for which the muons are the primary leptons produced by the interaction. For this kind of event, referred to as a {\it track-like event}, the interaction volume is built taking into account the muon maximum range in water and rock, evaluated at the highest simulated neutrino energy. The interaction volume is then a cylinder made of a layer of rock and a layer of seawater, with the separation surface coincident with the sea bottom (see Fig.\ \ref{fig:volume}). The volume radius is equal to the can radius plus the muon maximum range in water. The layer of rock has a height equal to the muon maximum range in this medium. If up-going events are not simulated the rock layer height is set to zero. The layer of sea water has a height equal to the can height plus the maximum muon range in water and it can not exceed the detector site depth. If down-going events are not simulated, the sea water layer height is set equal to that of the can.

Tau neutrinos interacting in CC do not induce long-track tau events at the present GENIE maximum valid energy of 5 TeV. However, muons could result as product of the tau decay.
For this reason when CC tau neutrino interaction events are simulated, the interaction volume is defined on the basis of the muon range as in the case of CC interacting muon neutrinos.

When the GENIE extension to very high energies (up to $10^{10}$ GeV) will be available \cite{Garcia:2019hze}, development of a more sophisticated simulation of long-track tau events will be required.

\section{Generation of neutrino events}
\label{sec:gen}

A flux driver class implementing the standard GENIE GFluxI class~\cite{Andreopoulos:2015wxa} has been developed within gSeaGen and used to configure the GENIE event generation driver.
The interface generates the neutrino energy and the track information needed to simulate the interaction. 

The gSeaGen flux driver is able to generate neutrinos coming from diffuse sources (e.g.\ atmospheric neutrinos) or from point-like and extended astrophysical sources.
The neutrino energy is always generated according to a power law spectrum. However, the code provides for each event a weight that can be used to re-weight the generated neutrino events to the chosen astrophysical or atmospheric neutrino spectrum.

The neutrino generation algorithm implemented in the gSeaGen flux driver is described in this section.

\subsection{Generation of the neutrino direction}
\label{sec:direction}

The generation of the neutrino direction depends on the kind of simulated neutrino flux. 
In case of diffuse fluxes, the neutrino arrival direction, defined by the $\theta$ and $\phi$ angles, is randomly extracted according to an isotropic distribution and the neutrino direction cosines $(v_x,v_y,v_z)$ are computed according to:
  
\begin{equation}
\left\{
\begin{array}{lll}
v_x&=&-\sin(\theta)\cdot\cos(\phi)\\ 
        &   & \\
v_y&=&-\sin(\theta)\cdot\sin(\phi)\\
        &   & \\
v_z&=&-\cos(\theta)
\end{array}.
\right. 
\label{eq:det2dir} 
\end{equation}

If neutrinos from an astrophysical point-like source are simulated, the direction is generated along the apparent source trajectory in the sky due to the Earth's rotation. The trajectory is calculated using the source equatorial coordinates (declination $\delta$ and right ascension RA) and the geographic coordinates of the detector, provided by the user.

The local sidereal time (LST) is uniformly extracted between 0 and 24 hours and the corresponding source hour angle, $\hbox{HA}=\hbox{LST}-\hbox{RA}$, is calculated. The code then computes the horizontal coordinates of the source, altitude $a$ and azimuth $A$, according to

\begin{equation}
\begin{array}{lll}
\sin(a)  & = & \sin(\delta)\cdot\sin(\lambda)+\cos(\delta)\cdot\cos(\lambda)\cdot\cos(\hbox{HA})\\ 
        &   & \\
\tan(A) & = & \displaystyle{\frac{\sin(\hbox{HA})}{\cos(\hbox{HA})\cdot\sin(\lambda)-\tan(\delta)\cdot\cos(\lambda)}} + 180^\circ  \\
\end{array},
\label{eq:eq2hor} 
\end{equation}
where $\lambda$ is the detector latitude\footnote{The difference between geographical and geocentric latitudes is not taken into account since the corrections in the coordinates are negligible with respect to the angular resolutions of neutrino telescopes of tenths of a degree. Corrections for parallax are also neglected since they are irrelevant for the distant astrophysical neutrino sources \cite{Meeus:1991:AA:532892}.}. Source position may be given also in galactic coordinates 
and the transformation to the equatorial reference system is done by gSeaGen. 

\begin{figure}[h]
\centering
\includegraphics[scale=0.5]{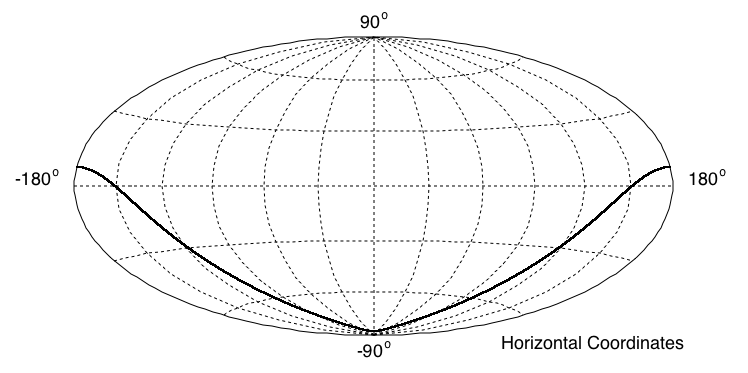}
\caption{Horizontal coordinates of neutrinos coming from the supernova remnant RX J1713.7-3946 direction generated with LST between 0 and 24 hours at the KM3NeT-ARCA site.}
\label{fig:rxj-all-horizontal}       
\end{figure}

As an example, neutrinos coming from the supernova remnant RX J1713.7-3946 direction  (RA:  17$^{\hbox{h}}$ 13$^{\hbox{m}}$ 33.1$^{\hbox{s}}$, $\delta$: $-39^\circ$ 45' 44'') have been simulated. The generated horizontal coordinates at the KM3NeT-ARCA site (latitude: N  36$^\circ$ 17' 48.34'', longitude: E 15$^\circ$ 58' 42.25'') are plotted in Fig. \ref{fig:rxj-all-horizontal}.

Once the source position in the sky is generated, the neutrino arrival direction angles are computed using Eq.\ \ref{eq:hor2det} and then transformed into the neutrino direction cosines with Eq. \ref{eq:det2dir}.

The user also has the possibility to generate the events in a given time interval, simulating neutrinos from transient sources. 
In this case the modified Julian day (MJD) is uniformly extracted between the two selected dates.
The value of the LST at the extracted MJD is then calculated to determine the horizontal coordinates of the source.

Finally, gSeaGen gives the possibility to simulate neutrinos coming from an extended astrophysical source with a small angular radius $R_s$ ($R_s\raisebox{-0.13cm}{~\shortstack{$<$ \\[-0.07cm]$\sim$}}~10^\circ$).
The generation proceeds as in the case of a point-like source with astronomical coordinates coincident with the extended source centre. 
Then, the neutrino directions are randomly generated according to a uniform distribution within a disk of radius $R_s$.

\begin{figure}[h]
\centering
\includegraphics[scale=0.5]{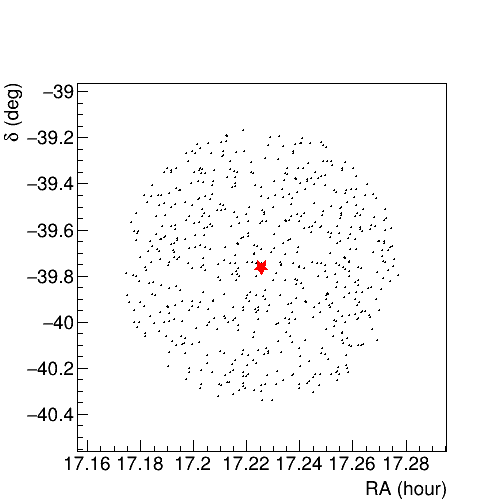}
\caption{Equatorial coordinates of neutrinos generated from the extended source RX J1713.7-3946 with $R_s=0.6^\circ$. The red star represents the coordinates of the source.}
\label{fig:rxj-ext-real}       
\end{figure}

For example, the arrival direction of neutrinos coming from RX J1713.7-3946 has been randomised in a circle of $R_s=0.6^\circ$, corresponding to the source extension. The equatorial coordinates of the generated neutrinos are plotted in Fig. \ref{fig:rxj-ext-real}.

All astronomical algorithms needed in the simulation of astrophysical sources have been implemented in a gSeaGen internal library based on Ref. \cite{Meeus:1991:AA:532892}.

\subsection{Generation of the neutrino vertex}
\label{sec:vertex}

\begin{figure}[ht]
\centering
\includegraphics[scale=0.45]{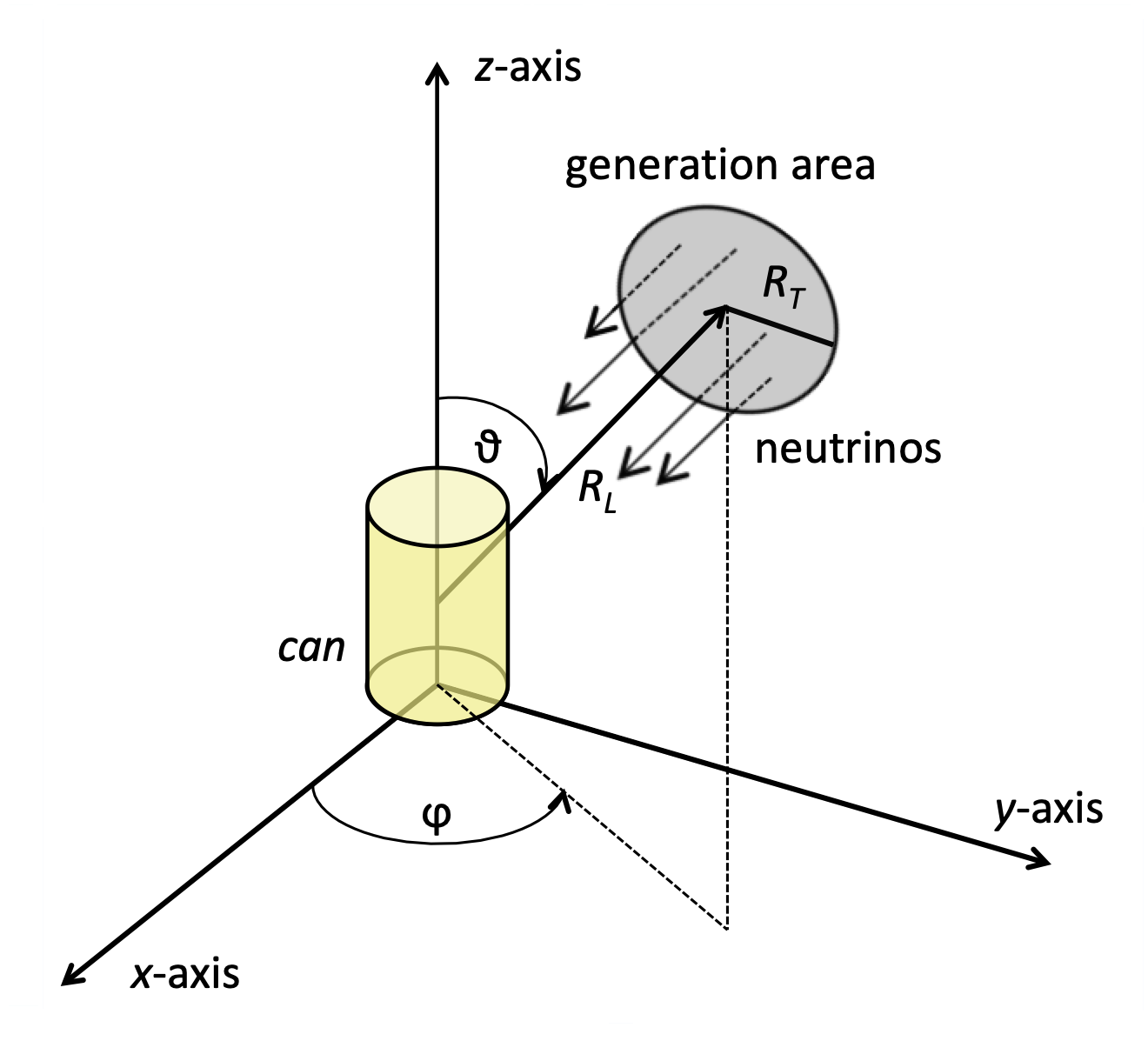}
\caption{Definition of the generation area for the neutrino vertices. For a given neutrino direction, the generation area is a circle, with radius $R_T$, which is tangent to a sphere of radius $R_L$ centred at the detector can centre. $R_T$ and $R_L$ are chosen such that the generation area is always outside the interaction volume and so that, for every given direction, the ``shadow'' of the generation surface covers the entire detector can. See text for more details.}
\label{fig:GenArea}       
\end{figure}

As in the standard GENIE neutrino flux drivers, the neutrino vertex is generated on a circular surface with radius $R_T$. The {\it generation area} is tangent to a sphere of radius $R_L$, centred at the detector can centre (see  Fig. \ref{fig:GenArea}). The radius $R_L$ is equal to the diagonal of the interaction volume so that the generation area is always outside this volume.
The radius of the generation area is: 

\begin{equation}
R_T= D/2 + 100 \hbox{ m},
\end{equation} 
where $D$ is the detector diagonal. Using option \texttt{-rt}, the user has the possibility to modify the radius of the generation area setting $D$ equal to the interaction volume diagonal or $R_T$ to a given value (see Sec. \ref{sec:usage}).

A circular generation area with a fixed radius disfavours particular directions when the detector is strongly asymmetric. 
Thus, an alternative new method has been developed to spread neutrino vertices within the surface defined by the projection of the can on to the plane perpendicular to the neutrino direction. This increases the number of events at the can 
by 30\% or more, depending on the can shape. When this method is activated (with option \texttt{-rt}), the generation area is not constant but depends on the neutrino direction.

\subsection{Generation of the neutrino energy}
\label{sec:energy}

Neutrinos are generated with energy following a power-law spectrum $E^{-X}$, where $X$ is the spectral index input by the user.
gSeaGen gives the possibility to bin the energy spectrum in equal divisions in $\log_{10}(E)$.  
The number $N^i$ of incoming neutrinos in each bin, ranging from $E_{\rm min}^i$ to $E_{\rm max}^i$, is given by:

\begin{equation}
 N^i=N_{\rm Tot}\cdot\frac{\int_{E_{\rm min}^i}^{E_{\rm max}^i}E^{-X}dE}
 {\int_{E_{\rm min}}^{E_{\rm max}}E^{-X}dE},
\end{equation} 
where $N_{\rm Tot}$ is the total number of neutrinos in the whole energy range from $E_{\rm min}$ to $E_{\rm max}$.
When muon inducing events are simulated, the interaction volume is scaled according to the maximum muon range in rock and water, calculated at the highest neutrino energy in the bin (see Sec. \ref{sec:vol}). 

Since interaction probabilities are very small numbers, GENIE scales up the probabilities to reduce the number of trials. This is reflected in the weights of interacting neutrinos as discussed in Sec.\ \ref{sec:wgt}.
The probability scale is defined as the maximum possible total interaction probability, i.e. the probability at the maximum neutrino energy (corresponding to the maximum cross section) and for the maximum possible neutrino path length in the interaction volume \cite{Andreopoulos:2009rq}.
When the energy range is binned, the interaction probability is scaled up at each bin, increasing the statistics for lower neutrino energies. In this way, the simulation can span different orders of magnitude in energy, as usually required in the simulation of neutrino telescopes.

\subsection{Neutrino interaction and events at the detector}
\label{sec:interaction}

Once the neutrino energy, position and direction are generated, the neutrino interaction is simulated using the GENIE event generation driver class GMCJDriver \cite{Andreopoulos:2009rq}.
If the neutrino interacts inside the can, all particles in the final state are stored in the output file. In the case of CC interaction of muon or tau neutrinos, the interaction could take place outside the can and still produce a visible lepton in the detector. In this case, all induced particles are rejected except the muons produced at a distance from the can lower than the maximum range in water. These muons are propagated and stored in the output file if they reach the can surface.

It is possible to add a layer of rock just below the can using the option \texttt{-bedrock} (see Sec. \ref{sec:usage}). In this case, all particles produced in the interaction are stored in the output file and the bedrock layer is considered part of the can for their propagation.

By default, gSeaGen uses its internal muon propagator code PropaMuon, described in \ref{sec:propamuon}. Alternatively, it is also possible to use external propagators such as MUSIC  \cite{Antonioli:1997qw} and PROPOSAL \cite{Koehne:2013gpa}, if enabled at compile-time.

\section{Computation of the event weight}
\label{sec:wgt}

\begin{figure}[h]
\centering
\includegraphics[scale=0.4]{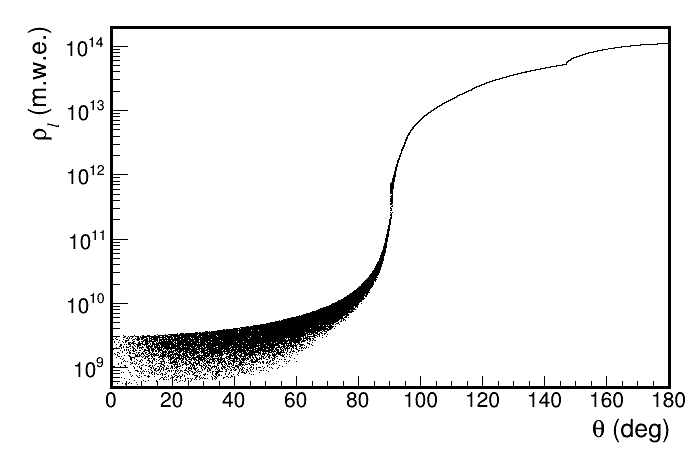}
\caption{Column depth as a function of the neutrino arrival direction according to the PREM model \cite{PREM}. The spread visible for downward going neutrinos is due to the range of interaction vertex depths from the same arrival direction.}
\label{fig:ColumnDepth}       
\end{figure}

\begin{figure}[ht]
\centering
\includegraphics[scale=0.4]{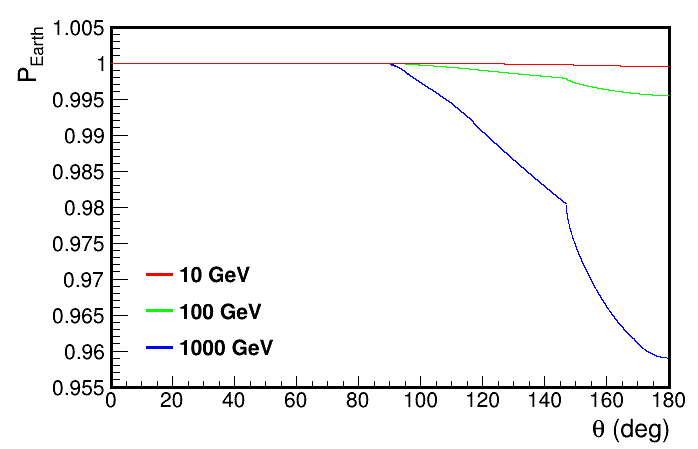}
\caption{Transmission probability through the Earth $P_{\rm Earth}$ as a function of the neutrino arrival direction and for different neutrino energies.}
\label{fig:PEarth}       
\end{figure}

During the simulation, each event is assigned a weight $w_{\rm evt}$ in order to shape the generation to the expected neutrino flux. The event weight is the product of the  generation weight $w_{\rm gen}$ and of the value of the neutrino energy spectrum $f(E,\theta,\phi)$: 
\begin{equation}
w_{\rm evt}=w_{\rm gen}\cdot f(E,\theta,\phi). 
\label{eq:wevt}
\end{equation}
The generation weight is:
\begin{equation}
w_{\rm gen}= \frac{I_E \cdot I_\theta \cdot T_{\rm gen} \cdot A_{\rm gen} \cdot N_\nu \cdot E^X \cdot P_{\rm scale}  \cdot P_{\rm Earth}(E,\cos\theta)}{N_{\rm Tot}}, 
\end{equation}
where:
\begin{itemize}

\item $I_E$ is the energy phase space factor, equal to the generation spectrum $E^{-X}$ integrated over the whole simulated neutrino energy range.

\item $I_\theta$ is the angular phase space factor, depending on the type of neutrino source:
\begin{equation}
I_\theta=\left\{
\begin{array}{lll} 
2\pi \cdot((\cos\theta)_{\rm max} -(\cos\theta)_{\rm min}) &  & \hbox{for diffuse flux (sr)} \\
                                                            &  & ~~~~~~~~~~~\\
 1                                                         &  & \hbox{for point/extended sources } \\
                                                           &  & \hbox{(dimensionless)} \\
\end{array}.
\right. 
\end{equation}

\item $T_{\rm gen}$ is the simulated livetime, equal to one year expressed in seconds. In the case of point-like or extended sources for which 
neutrinos are generated between two dates defined by the user (see Sec. \ref{sec:direction}), $T_{\rm gen}$ is set to the corresponding time interval.

\item $A_{\rm gen}$ is the area of the generation surface expressed in m$^2$. It has the constant value $\pi R_T^2$ when the generation area is defined as a circle with a fixed radius. 
When the user activates the generation on the can projection (see Sec. \ref{sec:vertex}), $A_{\rm gen}$ depends on the neutrino direction and is calculated for 
each event by the code.

\item $N_\nu$ is the number of generated neutrino types.

\item $E^{X}$ is the reciprocal of the generation spectrum evaluated at the generated neutrino energy.

\item $N_{\rm Tot}$ is the total number of simulated incoming neutrinos.

\item $P_{\rm scale}$ is the GENIE interaction probability scale \cite{Andreopoulos:2009rq} (see Sec. \ref{sec:energy}). 

\item $P_{\rm Earth}$ is the transmission probability through the Earth evaluated as:
\begin{equation}
P_{\rm Earth}(E,\cos\theta)=e^{\displaystyle{-N_A\cdot\sigma(E)\cdot\rho_l(\theta)}},
\end{equation}
where $\sigma(E)$ is the total CC cross section per nucleon (accounting for the different layer compositions). 
This ignores the (small) effect of NC interactions, which will reduce the neutrino energy without absorbing it.
The tau regeneration is also ignored since it is not relevant at the present neutrino energies and will be implemented in the future when the GENIE extension to very high energies will be available \cite{Garcia:2019hze}.
$\rho_l(\theta)$ is the column depth along the neutrino path inside the Earth up to the interaction vertex. It is computed as the line integral $\rho_l(\theta)=\int_L\rho_{\rm Earth}(r)dl$, where $L$ is the neutrino path at the angle $\theta$ and $\rho_{\rm Earth}(r)$ is the Earth density profile according to the PREM model \cite{PREM} (see Tab. \ref{tab:prem}). 
The column depth and the transmission probability for different neutrino energies calculated by gSeaGen are shown in Fig. \ref{fig:ColumnDepth} and \ref{fig:PEarth}, respectively. 

\end{itemize}

\begin{table}[h] 
\caption{Earth density profile according to the PREM model \cite{PREM}: the variable $\chi$ is the normalised radius $\chi=r/R_E$, where $r$ is the distance form the Earth centre and $R_E=6371$ km is the Earth radius. SiteDepth is the detector site depth. The third column indicates the target medium associated to each layer.} 
\begin{center} 
\begin{small} 
\begin{tabular}{lll} 
\hline \hline 
Density  (g/cm$^3$) & Radius (km) & Composition\\
\hline
$13.0885-8.8381 \chi^2$ 							& $0 < r< 1221.5$        			& Core    \\
$12.5815-1.2638 \chi-3.6426 \chi^2-5.5281 \chi^3$ 		& $1221.5 < r< 3480$  			& Core   \\
$7.9565-6.4761 \chi+5.5283 \chi^2-3.0807 \chi^3$ 		& $3480 < r< 5701$       			& Mantle  \\
$5.3197-1.4836 \chi$ 							& $5701 < r< 5771$      			& Mantle  \\
$11.2494-8.0298 \chi$ 							& $5771 < r< 5971$      			& Mantle  \\
$7.1089-3.8045 \chi$ 							& $5971 < r< 6151$      			& Mantle  \\
$2.691+0.6924 \chi$ 							& $6151 < r< 6346.6$    			& Mantle  \\
2.900 										& $6346.6 < r< 6356$    			& Mantle  \\
$\rho_{r}$ 										& $6356 < r< R_E-\hbox{SiteDepth}$      	& Rock    \\
$\rho_{s}$										& $R_E- \hbox{SiteDepth} < r< R_E$    & SeaWater\\
\hline \hline 
\end{tabular} 
\end{small} 
\end{center} 
\label{tab:prem}
\end{table}

\section{Calculation of the systematic weights}  
\label{sec:sys}

The known accuracy of the input simulation parameters can be propagated with GENIE to provide uncertainties related to neutrino interactions.
For each input physics quantity $P$, a systematic parameter $x_P$ is introduced. Tweaking it modifies the corresponding physics parameter $P$ as follows:
\begin{equation}
P\to P'= P(1 +x_P\cdot\delta P/P),
\end{equation}
where $\delta P$ is the estimated standard deviation of $P$. 
The calculation of the systematic errors in GENIE is based on an event reweighting strategy. A description of the full reweighting scheme is reported in \cite{Andreopoulos:2015wxa}. 

The evaluation of the systematics has been implemented in gSeaGen, using the GENIE class GReWeight \cite{Andreopoulos:2015wxa}. 
The implementation accepts single parameters or a list of them as input. In the latter case, the code treats all parameters at the same time and calculates the global systematic weight.  
If the calculation is activated, the systematic weights $w_{\rm sys}$ are written in the output file. The modified distributions are obtained by multiplying the event weights by $w_{\rm sys}$.

It is also possible to compute the systematic weights using the standard GENIE applications. These applications need as input the native GENIE format output file, available running gSeaGen with the option \texttt{-w} (see Sec. \ref{sec:usage}).






\section{Installing the code}

The gSeaGen source code can be downloaded from zenodo.org/record/3715310 \cite{gSeaGen}.
Its compiling requires a Linux-like operating system and some basic tools, such as the gcc compiler suite, make and PERL.
A minimal installation of gSeaGen requires only GENIE\footnote{The installation of GENIE is described in its user manual \cite{Andreopoulos:2015wxa} and in its web site http://www.genie-mc.org.} as external package.

A number of environmental variables must be defined:

\begin{itemize}
\item GSEAGEN  pointing at the top level gSeaGen directory;
\item GENIE pointing at the top level GENIE directory;
\item ROOTSYS  pointing at the top level ROOT directory.
\end{itemize}

gSeaGen has been written using the GENIE Auk production release series and tested with GENIE version 2.12.10. 
It is also possible to link gSeaGen with the new Bear release series (GENIE v3.0.0 and later). In this case the GENIE environmental variables
for compiling are:
\begin{itemize}
\item GENIE  pointing at the top level GENIE Generator directory;
\item GENIE\_REWEIGHT  pointing at the top level GENIE Reweight directory.
\end{itemize}

When the environmental variables are set, a minimal installation of gSeaGen is obtained typing:

\begin{verbatim}
$ cd $GSEAGEN
$ ./configure
$ make 
\end{verbatim}

The configuration script allows enabling/disabling features and specifying paths to external libraries such as the  MUSIC and PROPOSAL propagators.
The command
\begin{verbatim}
$ ./configure --help
\end{verbatim}
lists the configuration options.
The configuration script also creates the shell scripts setenv.sh and setenv.csh, helping the user to prepare the environment to run the code.

\section{Usage}
\label{sec:usage}

To set the environment configuration required to run gSeaGen, the setenv script must be executed: 
\begin{alltt}
$ source $GSEAGEN/setenv.csh {\normalfont (in tcsh or csh shells)}
$ source $GSEAGEN/setenv.sh {\normalfont (in bash shell)}
\end{alltt} 

The environmental variables required by GENIE and the ROOT version used during the compilation step are also set.
Finally, the user can run the code with the following syntax:

\begin{verbatim}
$ gSeaNuEvGen [OPTIONS]
\end{verbatim}

Running the code with the option \texttt{-h} (i.e. \texttt{\$ gSeaNuEvGen -h}), the code writes the list of the available options: 

\begin{alltt}
-a generation_spectral_index [default: 1.4]
   {\normalfont Specifies the generation spectral index.}

-b n_of_bins [default: 1]
   {\normalfont Specifies the number of energy bins in equal divisions in log\(\sb{10}(E)\).}

-bedrock bedrock_height [default: 0]
   {\normalfont Specifies the height of the bedrock just below the sea bottom (in meters).}

-c "CAN:Zmin,Zmax,Rad" [default: "CAN:0.000,659.862,309.450"]
   {\normalfont Specifies the detector can size where Zmin and Zmax are bottom and top Z 
     coordinates of the can and Rad is its radius in meters.}

--coord "UNIT:longitude,latitude" [default: RAD:0.74700,0.10763]
   {\normalfont Sets detector site latitude and longitude in radians (UNIT=RAD) or in 
      degrees (UNIT=DEG). The longitude is positive westwards from the 
      meridian of Greenwich, and negative to the East.} 

--cross-sections XSecName [default: gxspl-FNALsmall.xml]
   {\normalfont Inputs the name of an XML file with pre-computed cross-section 
      values used for constructing splines for GENIE\footnote{Pre-computed cross-sections for GENIE are available on http://www.genie-mc.org.}.} 

-d depth [default: 2475]
   {\normalfont Specifies the site depth in meters.}

-e min_energy,max_energy [default: 1,1000]
   {\normalfont Specifies the neutrino energy range in GeV.}

--event-generator-list list_name [default: Default]
   {\normalfont List of event generators to load. Valid settings are the XML block names appearing 
      in $GENIE/config/EventGeneratorListAssembler.xml.}

       eg. --event-generator-list  CC 
       eg. --event-generator-list  NC     
     
-f "simul:flux[nu_code],..."
   {\normalfont Specifies the types of neutrinos to be generated and the neutrino fluxes 
      used to weight the events (see Sec. \ref{sec:nu_flux} for details).}

-n n_of_neutrinos
   {\normalfont Specifies the number of incoming neutrinos.}

-o output_event_file_prefix [default: gseagen]
   {\normalfont Sets the prefix of the output event file. The output filename is built as: 
     [prefix].[run\_number].root.}

-point source_info
   {\normalfont Activates the generation from point-like and extended sources and 
      specifies the source information (see Sec. \ref{sec:source_info} for details).}

-prop prop_code [default: PropaMuon]
   {\normalfont Sets the muon propagation code. Possible choices are: PropaMuon 
      (internal code), MUSIC and PROPOSAL if enabled at compile-time.}

-r run_number [default: 100000000]
   {\normalfont Specifies the Monte Carlo run number.}

-rt [default: can]
   {\normalfont Set option to define the generation area radius \(R\sb{T}\). Possible choices are:
      1)}  -rt can 
   {\normalfont the generation surface will be a circle covering the detector can projected
      onto a plane perpendicular to each neutrino direction;
      2)}  -rt vol 
   {\normalfont the generation surface will be a circle covering the interaction 
      volume projected onto a plane perpendicular to each neutrino direction;
      3)}  -rt proj 
   {\normalfont the generation surface will be the detector can projected onto a plane;
      4)}  -rt value 
   {\normalfont the generation surface radius with \(R\sb{T}\) = value (in meters).}     

--seed random_number_seed [default: 12345]
   {\normalfont Specifies the random seed number.}

-t min_costheta,max_costheta [default: 0.,1. (up-going neutrinos)]
   {\normalfont Specifies the angular range of the neutrino arrival directions.}    

--tune generator_tune [default: G18_02a_00_000]
   {\normalfont Specifies the generator tune, required only for GENIE version }\(\ge\){\normalfont 3.00.00.}

-w        
   {\normalfont Writes the native GENIE output in [prefix].[run\_number].root}


\end{alltt}

\section{Input Files}

\subsection{Neutrino flux files}
\label{sec:nu_flux}

The user can define the type of neutrino and the neutrino fluxes used to weight the events with the option \texttt{-f} at run time. This does not affect the energy of the generated neutrinos since the power-law of the generation is controlled by options \texttt{-e} and \texttt{-a}. 
The general syntax of option \texttt{-f} is: 
\begin{alltt}
-f "simul:flux,...[nu\_code],flux...[nu\_code];simul:..."{\normalfont,}
\end{alltt}
where the tag simul specifies the flux model, flux is the name of the file containing the flux values and nu\_code is the neutrino type in the PDG code \cite{PhysRevD.98.030001} (12: $\nu_e$, $-12$: $\bar{\nu}_e$, 14: $\nu_\mu$, $-14$: $\bar{\nu}_\mu$, 16: $\nu_\tau$, $-16$: $\bar{\nu}_\tau$).

The neutrino flux can be provided with an ASCII file, containing the value of the neutrino spectrum at different neutrino energies and arrival directions. 
In order to get a continuously weighted flux, the code interpolates the tabulated values at the generated neutrino energy and direction.  
Possible values for the simul tag are BARTOL, FLUKA and HONDA that allow to read directly in the formats available on the web sites of Ref.s \cite{bartol,fluka,honda}, respectively. For example:

\begin{alltt}
-f "BARTOL:f210\_3\_z.kam\_num[14]"{\normalfont,}
-f "FLUKA:sk\_numu02.dat[14]"{\normalfont,}
-f "HONDA:grn-ally-20-12-solmin.d[14]"{\normalfont.}
\end{alltt}

Different neutrino types or neutrinos and anti-neutrinos can be simulated simultaneously, e.g.:

\begin{alltt}
-f "BARTOL:f210\_3\_z.kam\_num[14],f210\_3\_z.kam\_nbm[-14]"{\normalfont,}
-f "HONDA:grn-ally-20-12-solmin.d[14,-14]"{\normalfont.}
\end{alltt}

It is possible to define more than one file for each neutrino code. Fluxes from different files will be added, e.g.:

\begin{alltt}
-f "BARTOL:fmin10\_0401z.kam\_nue,f210\_3\_z.kam\_nue[12]"{\normalfont.}
\end{alltt}

The neutrino flux can also be provided using a user-defined function of the neutrino energy. In this case the tag simul has value FUNC1
and \texttt{flux} is a mathematical function (C++ syntax) depending on x, which represents the neutrino energy in units of GeV, and defining the neutrino flux in units of GeV$^{-1}$m$^{-2}$s$^{-1}$sr$^{-1}$, e.g.:

\begin{alltt}
-f "FUNC1:1E-5*pow(x,-2)[14]"{\normalfont.}
\end{alltt}
 
Two-dimensional functions depending on x and y can be provided to define a dependence on the neutrino arrival direction. In this case  the tag simul has value FUNC2 and the variable y represents $\cos(\theta)$. Similarly three-dimensional functions depending also on the variable z (representing the angle $\phi$ in radians) can be provided by setting the tag simul to FUNC3.  

Furthermore, it is possible to combine different neutrino fluxes by defining more than one tag simul. Fluxes corresponding to the same neutrino code will be combined, e.g.: 

\begin{alltt}
-f "BARTOL:f210_3_z.kam_num[14];FUNC1:1E-5*pow(x,-2)[14]"{\normalfont.} 
\end{alltt}

Finally, values for \texttt{simul} and \texttt{flux} are not mandatory, e.g.:
\begin{alltt}
-f "[14]"{\normalfont,}
-f "[12,14]"{\normalfont,}
-f "[16,-16]"{\normalfont.}
\end{alltt}
In this case the code sets by default \texttt{simul=FUNC1} and \texttt{flux=1E-5*pow(x,-2)}.
    
New fluxes will be implemented in the future to give the user the opportunity to simulate new sources of astrophysical interest. Regardless of the specified flux, it is possible to re-weight the generated sample with a different flux using Eq. \ref{eq:wevt}.

\subsection{Maximum muon range file}

In the case of muon and tau neutrinos undergoing CC interactions, the interaction volume is defined according to the maximum muon range in sea water and in rock as described in Sec. \ref{sec:vol}. 

By default, gSeaGen uses the muon ranges computed with MUSIC~\cite{Antonioli:1997qw} and contained in the file {\tt \$GSEAGEN/dat/muon\_rmax\_music.dat}.
Users can also provide their own file. The file should have three columns corresponding to the following items of information:
\begin{itemize}
\item    $\log_{10}(E)$, where $E$ is the muon energy in GeV,
\item    $\log_{10}(R_r)$, where $R_r$ is the range in rock in m.w.e.,
\item    $\log_{10}(R_w)$, where $R_w$ is the range in sea water in m.w.e.
\end{itemize}
The muon energy range in the file must be large enough to cover all possible energies of generated muons, while there are no constraints on the number of rows and on the energy binning. To define the new muon range input file, the user has then to define the environmental variable GMURNGFL pointing to the file with max muon ranges, e.g.:\\

\noindent GMURNGFL = GSEAGEN/dat/muon\_rmax\_music.dat.

\subsection{Target medium composition file}

To override default target medium compositions defined in Sec.\ \ref{sec:media}, the user has to define the environmental parameter MEDIACOMP, pointing to the file defining the new compositions. 
The input file must be in XML format where the root element must be media\_comp. The compositions are defined inside the tag param\_set. Its attribute media specifies the name of the target medium. 
Possible values for media are: SeaWater, Rock, Mantle and Core. Each chemical element is given through the tag param, where the attribute name indicates the chemical element name and the content defines the value of the mass fraction.
The list of elements from which the user can select to define a composition is: Al, Br, C, Ca, Cl, Fe, H, K, Mg, N, Na, Ni, O, S, Si, Ti. 
The mass number must not appear in the value of the attribute name, since it will be automatically set to the most abundant isotope.
For example, the file:
\begin{verbatim}
<?xml version="1.0" encoding="ISO-8859-1"?>
 <media_comp>
   <param_set media="SeaWater"> 
      <param name="O"> 0.8879 </param>
      <param name="H"> 0.1121 </param>
   </param_set>
   <param_set media="Rock" density="2.65">
      <param name="O">  0.533 </param>
      <param name="Si"> 0.467 </param>
   </param_set>
 </media_comp>
\end{verbatim}
will re-define the SeaWater and Rock compositions respectively with pure water and standard rock, while default compositions for Mantle and Core will be used.
The user has the possibility to set the density value (in g/cm$^3$) for SeaWater and Rock media, adding the attribute density to the tag param\_set. 
For example it is possible to use the SeaWater medium to simulate ice by writing:

\begin{verbatim}
<param_set media="SeaWater" density="0.92">.
\end{verbatim}

The attribute \texttt{density} is ignored for Mantle and Core media, for which the PREM density profile is always used.
For another example, the user can have a look at the file \$GSEAGEN/dat/MediaComposition.xml, which explicitly defines all the compositions with their default values. 

\subsection{Astrophysical source info file}
\label{sec:source_info} 

To activate the simulation of an astrophysical source, the user has to use the option \texttt{-point}, followed by information about the source. By default, a point-like source is simulated but an angular radius can also be set to simulate an extended circular source. It is possible to input just the source declination (\texttt{-point "DEC:Declination"}, with Declination in  degrees): the right ascension will be set to zero and the apparent point-like source trajectory will be simulated generating the LST between 0h and 24h. If point/extended mode is activated, only tag FUNC1 for the \texttt{-f} option is available in the present version.

The user also has the possibility to input a file containing information about the source (e.g.\ \texttt{-point "FILE:AstroSource.xml"}), such as the source equatorial or galactic coordinates, the date range in MJD and the source angular radius for an extended source.
The input file must be in XML format where the root element is astro\_source. Source information is set inside the tag param\_set, where its attribute source\_name specifies the name of the simulated source. Each piece of information is provided using the tag param, where the attribute name indicates the source parameter and the content defines the corresponding value. For example, giving the following file as input:

\begin{verbatim}
<?xml version="1.0" encoding="ISO-8859-1"?>
<astro_source>
 <param_set source_name="Dec-60">  
    <param name="Declination">  -60.   </param>
    <param name="RightAscension"> 0.   </param>
 </param_set> 
</astro_source>
\end{verbatim}
is equivalent to running the code with the option \texttt{-point "DEC:-60"}.
The complete list of all possible parameters is reported in Tab. \ref{tab:astro}. For example, the file \$GSEAGEN/dat/AstroSource.xml defines information for the star Antares and generates events during the whole of 2016.

\begin{table}[h] 
\caption{Complete list of input astronomical source parameters. If both equatorial and galactic coordinates are set, the equatorial ones will be used.
If both MJDs or dates are set, the MJDs will be used. If an interval time is set, via MJDs or dates, then TGen = (MJDStop-MJDStart)*86400 s, otherwise it will be TGen = 31556926 s (1 yr as in the diffuse mode).} 
\begin{center} 
\begin{small} 
\begin{tabular}{l|l} 
\hline \hline 
Parameter  & Description \\
\hline
Declination &	declination in deg\\
RightAscension &	right ascension in deg\\
GalacticLatitude &	galactic latitude in deg\\
GalacticLongitude &	galactic longitude in deg\\
Equinox &	equinox defining the galactic coordinates: J2000 or B1950\\
MJDStart 	& generation start time in Modified Julian Day\\
MJDStop 	& generation stop time in Modified Julian Day\\
DateStart 	& generation start date DD-MM-YYYY;HH:MM:SS \\
DateStop 	& generation stop date DD-MM-YYYY;HH:MM:SS\\
Radius 	& source angular radius in deg (point-source: Radius=0.)\\
\hline \hline 
\end{tabular} 
\end{small} 
\end{center} 
\label{tab:astro}
\end{table}

\subsection{Systematic parameter input file}

The calculation of the systematic errors related to the neutrino interaction models is activated by defining the environmental variable SYSTLIST. The variable must point at the input file containing the list of
physics parameters that can be varied and the associated systematic parameters (see Sec. \ref{sec:sys}).
The input file is in XML format where the root element must be syst\_param. The physics parameters are defined inside the tag param\_set which accepts the attribute set\_type. Setting set\_type=``list" defines a list of parameters. In this case, the code treats all parameters simultaneously and will calculate the global systematic weight. Each parameter to be considered for the calculation of the systematic weight is inserted with the tag param, where the attribute name indicates the physics parameter and the content defines the value of the corresponding systematic parameter, e.g.:

\begin{verbatim}
 <param_set set_type="list"> 
    <param name="MFP_pi"> +1.0 </param>
    <param name="MFP_N">  -1.0 </param>
 </param_set>.
\end{verbatim}

There is no limit to the number of lists that the user may define. The ``list" systematic weights will be output in the array WSys in the same order in which the lists are defined (see Sec. \ref{sec:output}).

Setting \texttt{set\_type="list;mirror"}, a second list with the same physics parameters and opposite systematic parameters is created. The two global weights will be output consecutively,  e.g.:

\begin{verbatim}
 <param_set set_type="list;mirror"> 
    <param name="MFP_pi">    +1.0        </param>
    <param name="MFP_N">     -1.0        </param>
 </param_set>
\end{verbatim}
is equivalent to writing:
\begin{verbatim}
 <param_set set_type="list"> 
    <param name="MFP_pi">    +1.0        </param>
    <param name="MFP_N">     -1.0        </param>
 </param_set>

 <param_set set_type="list"> 
    <param name="MFP_pi">    -1.0        </param>
    <param name="MFP_N">     +1.0        </param>
 </param_set>
\end{verbatim}

Setting \texttt{set\_type="parameter"}  defines single physics parameters. In this case, the code will calculate the systematic weights for each parameter, separately. The results will be output in dedicated tags: WSysGEN\_ParName, where ParName is the name of the systematic parameter.

As before, each physics parameter is input with the tag param, where the attribute name indicates the parameter name. In this case the content defines the number of the systematic parameter tweaking dial values between $-1$ and 1. For example, when setting: 

\begin{verbatim}
 <param_set set_type="parameter"> 
    <param name="MFP_pi">    5        </param>
    <param name="MFP_N">     3        </param>
 </param_set>
\end{verbatim}
the code will compute the systematic weights for the systematic parameter values $-1$, $-0.5$, 0, $+0.5$, $+1$ for the physics parameter MFP\_pi and for values $-1$, 0, $+1$ for MFP\_N.

An example can be found in the file \$GSEAGEN/dat/SystParams.xml. 
For the complete list of the physics parameters considered in GENIE, see table 9.1 of Ref. \cite{Andreopoulos:2015wxa}.

\section{Output File}
\label{sec:output}

The native gSeaGen event output is a ROOT file containing two ``trees": Header and Events.

The Header tree reports the general information about the simulation: 

\begin{itemize}
\item {\bf gSeaGenVer}  ({\it string}): gSeaGen version;
\item {\bf GenieVer}  ({\it string}): GENIE version;
\item {\bf RunTime}  ({\it string}): running time;
\item {\bf RunNu} ({\it int}): run number;
\item {\bf RanSeed} ({\it int}): random generator seed;
\item {\bf EventGeneratorList} ({\it string}): list of simulated interaction channels;
\item {\bf InpXSecFile}  ({\it string}): cross-section spline file;
\item {\bf NTot}  ({\it double}): total number of generated incoming neutrinos;
\item {\bf EvMin, EvMax}  ({\it double}): generated neutrino energy range (GeV);
\item {\bf CtMin, CtMax}  ({\it double}): generated neutrino arrival $\cos\theta$ range;
\item {\bf Alpha}  ({\it double}): generation spectral index;
\item {\bf NBin} ({\it int}): 	number of energy bins;
\item {\bf Can1, Can2, Can3}  ({\it double}): $Z^{\rm min}_{\rm can}$, $Z^{\rm max}_{\rm can}$ and $R_{\rm can}$ of the can (m);
\item {\bf HRock, HSeaWater, RVol}  ({\it double}): interaction volume: height in rock and water and radius (m);
\item {\bf SiteDepth}  ({\it double}): 	site depth (m);
\item {\bf SiteLatitude, SiteLongitude}  ({\it double}): detector latitude and longitude (rad);
\item {\bf Agen}  ({\it double}): 	area of the generation surface (m$^2$);
\item {\bf RhoSW, RhoSR}  ({\it double}): SeaWater and Rock densities (g/cm$^3$);
\item {\bf TGen}  ({\it double}): generation time (s);
\item {\bf PropCode}  ({\it string}): muon propagation code;
\item {\bf GenMode}  ({\it string}): simulated livetime mode (DIFFUSE or POINT);
\item {\bf SourceFile}  ({\it string}): input file name with source information (POINT mode);
\item {\bf SourceName}  ({\it string}): name of the simulated source (POINT mode);
\item {\bf Declination, RightAscension}  ({\it double}): source declination and right ascension (rad) (POINT mode);
\item {\bf SourceRadius} ({\it double}): source angular radius (rad) (POINT mode);
\item {\bf MJDStart, MJDStop}  ({\it double}): source simulated interval in MJD (POINT mode only).
\end{itemize}

The Events tree reports the information about the generated events:

\begin{itemize}
\item {\bf Evt} ({\it int}): event Id number (incremented sequentially);
\item {\bf PScale}  ({\it double}): GENIE interaction probability scale (see Sec. \ref{sec:energy});
\item {\bf TargetA, TargetZ}  ({\it int}): mass number and atomic number of the target nucleus; 
\item {\bf InterId} ({\it int}): interaction type Id, according to GENIE enumeration;
\item {\bf ScattId}  ({\it int}): scattering type Id, according to GENIE enumeration;
\item {\bf Bx, By}   ({\it double}): Bjorken kinematic variables x and y;
\item {\bf LST}  ({\it double}): Local Sidereal Time (rad) (POINT mode);
\item {\bf MJD}   ({\it double}): 	Modified Julian Day, if generated (POINT mode);
\item {\bf VerInCan}  ({\it int}): 1 if the interaction vertex is inside the can, 2 if it is inside the bedrock and 0 outside. 
\item {\bf WaterXSec}  ({\it double}): neutrino cross section per nucleon in pure water (m$^2$);
\item {\bf WaterIntLen}  ({\it double}): neutrino interaction length in pure water (m);
\item {\bf PEarth}  ({\it double}): transmission probability through the Earth;
\item {\bf ColumnDepth}  ({\it double}): column depth, entering in the calculation of PEarth (m.w.e.);
\item {\bf XSecMean}   ({\it double}): average cross section per nucleon (m$^2$) along the neutrino path, to be considered in the calculation of PEarth;
\item {\bf GenWeight}  ({\it double}): generation weight (GeV m$^2$ s sr for diffuse fluxes and GeV m$^2$ s for point-like/extended sources);
\item {\bf EvtWeight}  ({\it double}): 	event weight;
\item {\bf E\_nu}  ({\it double}):  neutrino energy (GeV);	
\item {\bf Pdg\_nu}  ({\it int}): 	neutrino PDG code;
\item {\bf Vx\_nu, Vy\_nu, Vz\_nu}  ({\it double}): coordinates of interaction vertex in meters;
\item {\bf Dx\_nu, Dy\_nu, Dz\_nu}  ({\it double}): ceutrino track direction cosines;
\item {\bf T\_nu}  ({\it double}): time of neutrino interaction (ns); 
\item {\bf E\_pl}  ({\it double}):  primary lepton energy (GeV);	
\item {\bf Pdg\_pl}  ({\it int}): primary lepton PDG code;
\item {\bf Vx\_pl, Vy\_pl, Vz\_pl}  ({\it double}): coordinates of interaction vertex (m);
\item {\bf Dx\_pl, Dy\_pl, Dz\_pl}  ({\it double}): primary lepton track direction cosines;
\item {\bf T\_pl}  ({\it double}): time of neutrino interaction (ns); 
\item {\bf NTracks}  ({\it int}): final number of particles in the event; 
\item {\bf Id\_tr}  ({\it int[NTracks]}): particle Id number; 
\item {\bf E\_tr}  ({\it double[NTracks]}): particle energy (GeV);	
\item {\bf Pdg\_tr}  ({\it int[NTracks]}): particle PDG code;
\item {\bf Vx\_tr, Vy\_tr, Vz\_tr}  ({\it double[NTracks]}): coordinates of final particle position (m);
\item {\bf Dx\_tr, Dy\_tr, Dz\_tr}  ({\it double[NTracks]}): particle track direction cosines;
\item {\bf T\_tr}  ({\it double[NTracks]}): time of passage through the final position with respect to the interaction time (ns); 
\item {\bf NSysWgt} ({\it int}): number of lists of systematic parameters;
\item {\bf WSys} ({\it double[NSysWgt]}): systematic weights for each list of parameters, in the same order in which the lists are defined;
\item {\bf NSysWgt\_ParName} ({\it int}): number of the tweaking dial values between $-1$ and 1 for the parameter ParName;
\item {\bf WSys\_ParName} ({\it double[NSysWgt\_ParName]}): systematic weights for each tweaking dial value for the parameter ParName.
\end{itemize}

\section{Applications}
\label{sec:res}

The gSeaGen code has been developed to provide the possibility to use maintained neutrino interaction codes/libraries.
The code has been used as a cross-check for GENHEN \cite{genhen}, the reference neutrino simulation code for the ANTARES experiment \cite{Collaboration:2011nsa}, written in FORTRAN, and presently used also by the KM3NeT Collaboration for the ARCA detector \cite{km3net-loi} simulation.

\begin{figure}[h]
\centering
\includegraphics[scale=0.37]{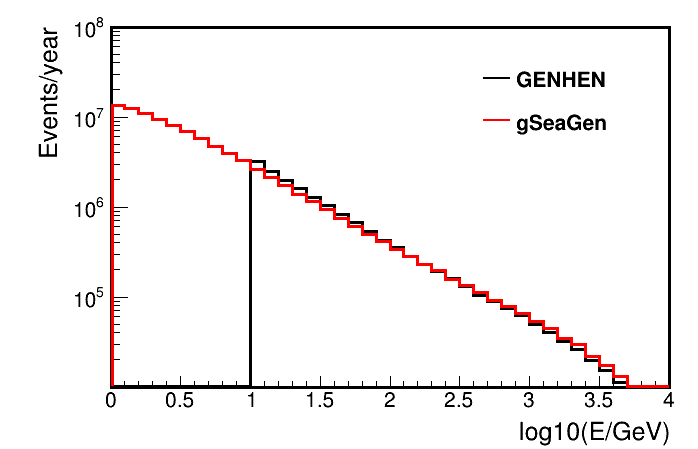}
\includegraphics[scale=0.37]{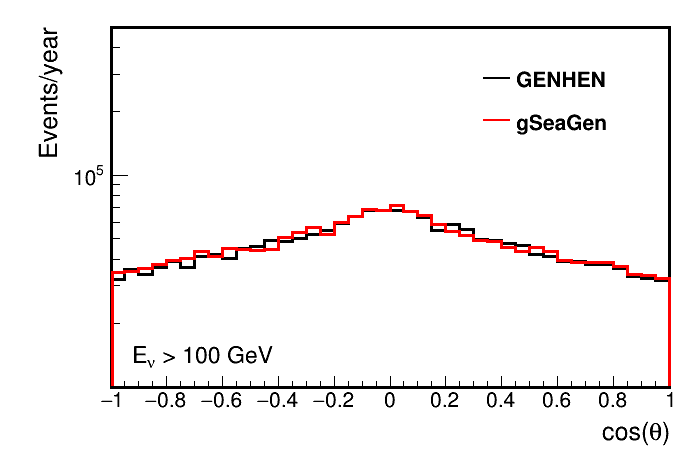}
\caption{Energy spectrum (top) and angular distribution (bottom) of atmospheric muon neutrinos producing detectable events. Generated events are weighted according to the Bartol atmospheric flux \cite{Barr:2004br}. The simulation was carried out with gSeaGen v6r1 \cite{gSeaGen} using GENIE version 2.12.10 \cite{GenieWeb}. Results from the standard neutrino event generator GENHEN  \cite{genhen} are shown for comparison. Note the difference in events per year in the 10-100 GeV range, due to GENHEN being optimised for events with energy above 100 GeV.}
\label{fig:muons}       
\end{figure}

As an example, muon neutrinos have been generated up to the present GENIE maximum valid energy of 5 TeV. The spectrum of generated events has been weighted according to the Bartol atmospheric muon flux model \cite{Barr:2004br}. The results are shown in Fig. \ref{fig:muons}.

GENHEN is not optimised at energies below $\sim 100$ GeV and can not be used to simulate interactions below 10 GeV. Moreover, it does not allow for the evaluation of systematics due to neutrino interaction models.
For these reasons, the gSeaGen code was used by the ANTARES Collaboration in the final data analysis to measure atmospheric neutrino oscillation parameters \cite{antares-oscillation}.

These limitations of GENHEN at lower energies are especially important for the KM3NeT Collaboration, hence gSeaGen has been chosen as the reference code for the simulation of the ORCA detector which is optimised to detect neutrinos of a few GeV and is aiming at measuring the neutrino mass hierarchy. The gSeaGen code was used to optimise the detector geometry and to estimate the expected performances published in the KM3NeT Letter of Intent \cite{km3net-loi}. At present, the code is  used to carry out the Monte Carlo simulations for the ORCA data analysis \cite{orca-icrc19}. 


Furthermore, both DeepCore and Upgrade detectors, which are the existing and the future infills of the existing IceCube neutrino telescope \cite{icecube-upgrade}, also aim at measuring neutrino oscillations in the GeV range. These detectors are located in glacial Antarctic ice and sit above a layer of bedrock. While the code default values and naming convention specify an in-water detector, gSeaGen will also be suitable for modelling these experiments, with an adaptation of medium properties. 

It is planned to extend the use of gSeaGen to the KM3NeT-ARCA and ANTARES high energy analyses when the GENIE extension at the PeV scale will be available \cite{Garcia:2019hze}. Finally, gSeaGen can be used for GVD \cite{gvd}, a next-generation neutrino telescope being constructed in Lake Baikal. gSeaGen should be well-suited to generate interactions at energies from a few TeV up to 100 PeV for GVD.

\section{Conclusions}

The gSeaGen code, written in C++ and based on the GENIE neutrino interaction code, has been developed within the KM3NeT Collaboration. The gSeaGen code generates all flavour neutrino interaction events detectable in a Cherenkov neutrino telescope. It can simulate neutrinos coming from diffuse sources (e.g.\ atmospheric neutrinos) and from point-like or extended astrophysical sources, according to several neutrino fluxes.

The gSeaGen code provides the possibility of changing the composition and the density of the neutrino target medium surrounding the detector, allowing the study of systematic uncertainties due to medium composition and also considering both under-water and under-ice detectors.

Presently, gSeaGen is the reference code for the simulation of the KM3NeT-ORCA detector, aiming at the study of neutrino physics and optimised to detect neutrinos in the GeV energy range. gSeaGen will be also used in the KM3NeT-ARCA and ANTARES TeV-PeV neutrino searches once the extension of the GENIE code up to the PeV scale will be available.

\appendix
\section{The PropaMuon code}
\label{sec:propamuon}

The PropaMuon library is a three-dimensional Monte Carlo code for muon propagation through a generic medium. The code simulates the energy losses and the angular deviations of muons taking into account ionisation and radiative processes (bremsstrahlung, direct electron pair production and nuclear interactions). 

The muon energy losses are expressed as the sum of the ionisation and radiative losses:
\begin{equation}
\frac{dE}{dx}=\left( \frac{dE}{dx}\right)_{\rm ion}  + \left( \frac{dE}{dx}\right)_{\rm rad}.  
\end{equation}

The fluctuations associated with the ionisation are small and the energy loss mechanism is treated as continuous. The mean value is evaluated with the Bethe-Bloch formula \cite{PhysRevD.98.030001}.
PropaMuon can consider any medium composition if ionisation coefficients are provided for each element of which it is made. At present only elements listed in Tab. \ref{tab:media} are considered.

The radiative energy loss is subdivided into a ``soft" part, treated as continuous, if the energy fraction $v$ transferred from the muon to the photon is below a certain threshold ($v<v_{\rm cut}$, here $v_{\rm cut}=10^{-3}$), and a ``hard" part ($v\ge v_{\rm cut}$) treated stochastically \cite{PhysRevD.44.3543}. Given an element with atomic mass A, present in the mixture, the radiative energy loss is:

\begin{equation}
\begin{array}{lll}
\displaystyle{\left( \frac{dE}{dx}\right)_{\rm rad} }& =\displaystyle{\left( \frac{dE}{dx}\right)_{\rm soft}  + \left( \frac{dE}{dx}\right)_{\rm hard} }\\
 & \\
                                               & = \displaystyle{\frac{N_A}{A} E \int_{0}^{v_{\rm cut}} {\frac{d\sigma(v,E)}{dv}dv} + \frac{N_A}{A} E \int_{v_{\rm cut}}^{1} {\frac{d\sigma(v,E)}{dv}dv}}.\\
\end{array}
\end{equation}

The total energy loss in the medium is evaluated as:

\begin{equation}
\frac{dE}{dx}=\sum w_i\cdot \left(\frac{dE}{dx}\right)_i,
\end{equation}
where $(dE/dx)_i$ is the energy loss in the $i$th element having a weight fraction  $w_i$.

Total cross section, mean free path and probability for ``hard radiation'' are:

\begin{equation}
\sigma_{\rm hard}(E)=\int_{v_{\rm cut}}^{1} {\frac{d\sigma(v,E)}{dv}dv},
\end{equation}

\begin{equation}
\lambda_{\rm hard}(E)=\frac{A}{\sigma_{\rm hard}(E)N_A},
\end{equation}

\begin{equation}
P_{\rm hard}(E)=\frac{dx}{\lambda_{\rm hard}(E)}.
\end{equation}

Differential cross sections for radiative processes are computed using the Petrukhin and Shestakov formula \cite{brem} for bremsstrahlung, the Kokoulin and Petrukhin formula \cite{pair} for direct electron pair production and the Bezrukov and Bugaev formula for the nuclear interaction \cite{nucl}.

Given a muon with energy $E$, the code extracts the hard radiation interaction path according to:

\begin{equation}
P(r)=\frac{1}{\lambda_{\rm hard}}\exp{(-r/\lambda_{\rm hard})}
\end{equation}
and selects the radiative process considering that:

\begin{equation}
\lambda_{\rm hard} = \lambda_{\rm hard}^{\rm brem} + \lambda_{\rm hard}^{\rm pair} + \lambda_{\rm hard}^{\rm nucl}.
\end{equation}

Then, the code computes the continuous energy loss (ionisation + soft radiation) and simulates the multiple scattering up to the interaction point. For the multiple scattering, the Lynch and Dahl formula \cite{mult} is used.

At this point, the code extracts the fraction $v$ of the energy lost by the muon in the selected ``hard radiation'' process from a numerical solution of the equation:

\begin{equation}
P= \frac{\displaystyle{\int_v^1 {\frac{d\sigma}{dv}dv}} }{ \displaystyle{\int_{v_{\rm cut}}^1 {\frac{d\sigma}{dv}dv}}}.    
\end{equation}

where $P$ is a random number between 0 and 1. The muon angular deviation due to the selected ``hard radiation'' process is simulated following the approximations proposed in Ref. \cite{VANGINNEKEN1995213}.
The entire procedure is then repeated until the muon is at rest or input depth is reached. 

\begin{figure}[ht]
\centering
\includegraphics[scale=0.4]{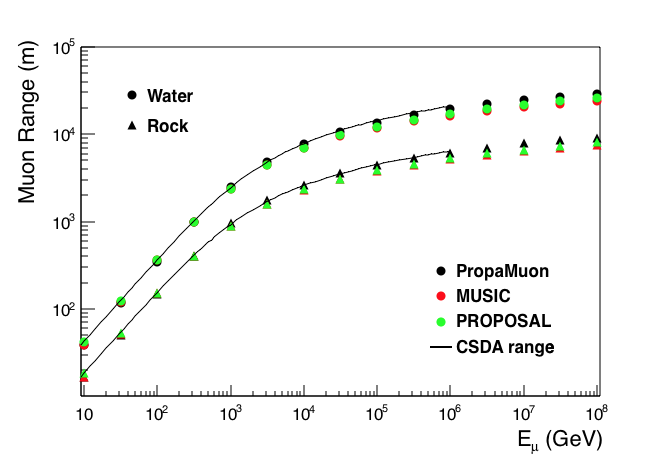}
\caption{Mean muon range simulated in water and standard rock with PropaMuon and for comparison with MUSIC and PROPOSAL propagators, linkable to gSeaGen. The CSDA range is also shown (values taken from Ref. \cite{ioncoeff}).}
\label{fig:RangePropagators}       
\end{figure}

The muon range simulated with the PropaMuon code in pure water and in standard rock is shown in Fig. \ref{fig:RangePropagators}. For comparison, the ranges simulated with MUSIC and PROPOSAL and the Continuous Slowing Down Approximation (CSDA)  range \cite{ioncoeff} are also shown.

\section*{Acknowledgements}
The authors acknowledge the financial support of the funding agencies:
Agence Nationale de la Recherche (contract ANR-15-CE31-0020),
Centre National de la Recherche Scientifique (CNRS), 
Commission Europ\'eenne (FEDER fund and Marie Curie Program),
Institut Universitaire de France (IUF),
IdEx program and UnivEarthS Labex program at Sorbonne Paris Cit\'e (ANR-10-LABX-0023 and ANR-11-IDEX-0005-02),
Paris \^Ile-de-France Region,
France;
Shota Rustaveli National Science Foundation of Georgia (SRNSFG, FR-18-1268),
Georgia;
Deutsche Forschungsgemeinschaft (DFG),
Germany;
The General Secretariat of Research and Technology (GSRT),
Greece;
Istituto Nazionale di Fisica Nucleare (INFN),
Ministero dell'Istruzione, dell'Universit\`a e della Ricerca (MIUR),
PRIN 2017 program (Grant NAT-NET 2017W4HA7S)
Italy;
Ministry of Higher Education, Scientific Research and Professional Training,
Morocco;
Nederlandse organisatie voor Wetenschappelijk Onderzoek (NWO),
the Netherlands;
The National Science Centre, Poland (2015/18/E/ST2/00758);
National Authority for Scientific Research (ANCS),
Romania;
Ministerio de Ciencia, Innovaci\'{o}n, Investigaci\'{o}n y Universidades (MCIU): Programa Estatal de Generaci\'{o}n de Conocimiento (refs. PGC2018-096663-B-C41, -A-C42, -B-C43, -B-C44) (MCIU/FEDER), Severo Ochoa Centre of Excellence and MultiDark Consolider (MCIU), Junta de Andaluc\'{i}a (ref. SOMM17/6104/UGR), Generalitat Valenciana: Grisol\'{i}a (ref. GRISOLIA/2018/119) and GenT (ref. CIDEGENT/2018/034) programs, La Caixa Foundation (ref. LCF/BQ/IN17/11620019), EU: MSC program (ref. 713673),
Spain.

\bibliographystyle{elsarticle-num}
\bibliography{biblio}

\end{document}